\documentclass{article}
\usepackage[utf8]{inputenc}
\usepackage{amsmath}
\usepackage{amsfonts}
\usepackage{amssymb}
\usepackage{amsthm}

\def\p{\partial}

\theoremstyle{remark}

\newcommand{\wt}{\widetilde}
\newcommand{\be}{\begin{equation}}
\newcommand{\ee}{\end{equation}}
\newcommand{\bea}{\begin{eqnarray}}
\newcommand{\eea}{\end{eqnarray}}
\newcommand{\beaa}{\begin{eqnarray*}}
\newcommand{\eeaa}{\end{eqnarray*}}

\newcommand{\nn}{\nonumber}

\newcommand{\diverg}{\mathop{\mathrm{div}}\nolimits}
\usepackage{authblk}
\author{L.V. Bogdanov
}
\affil{Landau Institute for Theoretical Physics RAS,
Chernogolovka}

\title{Dispersionless BKP equation, the Manakov-Santini system and Einstein-Weyl structures}
\date{}

\begin{document}

\maketitle

\begin{abstract}
We construct a map of solutions of the dispersionless BKP (dBKP) equation
to solutions of the Manakov-Santini (MS) system. This map defines
an Einstein-Weyl structure  corresponding to the dBKP equation through
the general Lorentzian Einstein-Weyl structure corresponding to the
MS system. We give a spectral characterisation of reduction
of the MS system which singles out the image of  the dBKP equation solutions
and also consider more general reductions of this class. 
We define the BMS system and extend the map defined above to the map
(Miura transformation)
of solutions of the BMS system to solutions of the MS system,
thus obtaining an Einstein-Weyl structure for the BMS system.
\end{abstract}
\section{Introduction}
Dispersionless BKP hierarchy is a reduction of the dispersionless KP hierarchy by a special symmetry,
which is compatible only with odd times of the hierarchy \cite{Takasaki}, \cite{BK2005}.
Equations of the hierarchy can be represented as compatibility conditions for the Hamilton-Jacobi  equations.
For the first equation of the hierarchy (dispersionless BKP equation)
corresponding Hamilton-Jacobi  equations are
\bea
&&
S_y=H_1=p^3 + 3up
\nn\\
&&
S_t=H_2=p^5+5u p^3+vp,  \quad p=S_x
\label{HJ0}
\eea
The symmetry characterising the reduction is a simple condition for Hamiltonians
$H(-p)=-H(p)$,
$x=t_1$, $y=t_3$, $t=t_5$ (in terms of dispersionless KP hierarchy times).
Compatibility condition of Hamilton-Jacobi equations (\ref{HJ0})
can be written in the form
\bea
\p_t H_1 - \p_y H_2 + \{H_1,H_2\}=0,
\eea
where the Poisson bracket is
$\{f,g\}=f_p g_x - f_x g_p$,
it gives the 
dispersionless BKP equation (see \cite{Takasaki}, \cite{BK2005})
\bea
\textstyle
\frac{1}{5}u_t+u^2u_x-
\frac{1}{3}uu_y-
\frac{1}{3}u_x\partial_x^{-1}u_y-
\frac{1}{9}\partial_x^{-1}u_{yy}=0.
\label{dBKP0}
\eea
In what follows we will rescale the times to simplify the coefficients of equation
and use the Hamiltonians
$H_1=\frac{1}{3}p^3 + up$,
$H_2=\frac{1}{5}p^5+ u p^3+vp$, then equation (\ref{dBKP0}) reads
\be
u_t+u^2u_x-
uu_y-
u_x\partial_x^{-1}u_y-
\partial_x^{-1}u_{yy}=0.
\label{dBKP}
\ee
In potential form, $u=f_x$, we have
\be
\textstyle
\partial_x\left(
f_{t}+ \frac{1}{3}f_x^3 -
f_x f_{y}
\right)
=
f_{yy}.
\label{pdBKP}
\ee

The Lax pair in terms of Hamilton-Jacobi equations
(pseudopotentials) can be represented as 
commutation relations for Hamiltonian vector fields
\bea
&&
V_1=\p_y - \{H_1,\dots\},
\nn\\
&&
V_2=\p_t - \{H_2,\dots\},
\label{Vham}
\eea
where $\{H,\dots\}=(\p_p H)\p_x - (\p_x H)\p_p$.
In this Lax pair $p$ plays a role of `spectral parameter',
and commutation relation $[V_1, V_2]=0$ gives exactly
equation (\ref{dBKP}). The formalism of integration 
of equations arising as commutation relations for vector
fields is not restricted to Hamiltonian vector fields.
Moreover, several interesting examples corresponding
to general vector fields were discussed, e.g.,
the Manakov-Santini (MS) system \cite{MS06}, \cite{MS07}, which was recently demonstrated
to describe a general local form of the Einstein-Weyl equations \cite{DFK14}.
\section*{From the dBKP equation to the MS system}
 In explicit form vector fields  (\ref{Vham}) read
\bea
&&
V_1=\p_y - (p^2+f_x) \p_x + f_{xx}p\p_p,
\nn\\
&&
V_2=\p_t - (p^4 + 3 f_{x}p^2 + v) \p_x,
+( f_{xx}p^2 + v_x)p\p_p,
\label{vfdBKP}
\eea
$v=f_y + f_x^2$.
Symmetry of vector fields $V(-p)=V(p)$ characterises
dispersionless BKP hierarchy in the framework of dKP hierarchy.

Let us transform the spectral variable
$p^2=\mu$, commutation relations evidently remain the same,
\beaa
&&
V_1=\p_y - (\mu + f_x) \p_x + 2f_{xx}\mu\p_\mu,
\\
&&
V_2=\p_t - (\mu^2 + 3 f_{x}\mu + v) \p_x
+2( f_{xx}\mu + v_x)\mu\p_\mu
\eeaa
Vector fields are still Hamiltonian (the bracket should 
be changed).

Now let us make a change
$\lambda=\mu+2f_x$ (which also preserves commutation relations)

\bea
&&
V_1=\p_y - (\lambda - f_x) \p_x + 2(f_{xy} - f_x f_{xx})
\p_\lambda,
\nn
\\
&&
V_2=
\p_t - (\lambda^2 - f_{x}\lambda + f_y
-f_x^2) \p_x
\nn
\\
&&\qquad\quad
+2((f_{xy}- f_{x}f_{xx})\lambda
 + 
(f_{xt} - f_y f_{xx} + f_x^2 f_{xx} - 2 f_{xy}f_x
))\p_\lambda. 
\label{BLax}
\eea
Lax pair $V_1$, $V_2$ has the structure of 
the Manakov-Santini (MS) system Lax pair
\be
\begin{split}
X_1&=\partial_y-(\lambda-v_{x})\partial_x + u_{x}\partial_\lambda,
\\
X_2&=\partial_t-(\lambda^2-v_{x}\lambda+u -v_{y})\partial_x
+(u_{x}\lambda+u_{y})\partial_\lambda,
\end{split}
\label{MSLax}
\ee
the MS system reads
\be
\begin{split}
u_{xt} &= u_{yy}+(uu_x)_x+v_xu_{xy}-u_{xx}v_y,
\\
v_{xt} &= v_{yy}+uv_{xx}+v_xv_{xy}-v_{xx}v_y
\end{split}
\label{MSeq}
\ee
A comparison of Lax pairs (\ref{MSLax}),
(\ref{BLax}) gives a map from solution of
dBKP equation (\ref{pdBKP}) to solution of  MS system (\ref{MSeq}),
\bea
v=f,\quad u = 2f_y -f_x^2,
\label{map}
\eea
corresponding solutions  of the  MS system
satisfy a reduction
\bea
u = 2v_y -v_x^2.
\label{red}
\eea
This map defines the Einstein-Weyl structure
corresponding to dBKP equation.
\subsection*{Einstein-Weyl structure for the dBKP equation}
Let us remind (see 
\cite{PT93}, \cite{FK14}, for more detail), 
that a Weyl space is a manifold with
a conformal structure
$[g]$ and a symmetric connection
${{D}}$ consistent with $[g]$ in a sense that
for every $g\in [g]$
\bea
{{D}}g=\omega\otimes g
\label{Weyl}
\eea
for some differential form $\omega$ (the connection
preserves a conformal class). 
Einstein-Weyl spaces are defined by the condition that
the trace-free part of the symmetrised Ricci tensor of
the connection  
${{D}}$ vanishes (Einstein equations),
which together with relation 
(\ref{Weyl}) 
constitute Einstein-Weyl equations system,
in a coordinate form
\bea
D_i g_{ij}=\omega_i g_{ij},
\quad R_{(ij)}=\Lambda g_{ij},
\label{EW0}
\eea
here $\Lambda$ is some function.
Einstein-Weyl equations are correctly
defined for arbitrary manifold dimension not
less than three,  but the most interesting case
is three-dimensional when they are integrable 
\cite{Hitchin}. 
The Manakov-Santini system (\ref{MSeq}) defines a general local form
of  the (2+1)-dimensional  Lorentzian Einstein-Weyl structure  (modulo 
coordinate transformations) with the metric $g$ and one-form (covector)
$\omega$ defined as \cite{DFK14}
\begin{gather}
\begin{split}
g &= -(dy + v_x dt )^2 +4(dx + (u - v_y ) dt ) dt,
\\
\omega &= v_{xx} dy+(-4u_x + 2v_{xy} +v_xv_{xx})dt,
\end{split}
\label{metricMS}
\end{gather}
where $u$, $v$ satisfy the MS system.
Using the map (\ref{map}), we obtain the Einstein-Weyl structure corresponding
to solutions of the potential dispersionless BKP equation (\ref{pdBKP}),
\begin{gather}
\begin{split}
g &= -(dy + f_x dt )^2 +4(dx + (f_y - f_x^2) dt ) dt,
\\
\omega &= f_{xx} dy+3(- 2f_{xy} + 3f_xf_{xx})dt,
\end{split}
\label{metricBKP}
\end{gather}
It could be possible to construct this Einstein-Weyl structure 
by the methods of 
the work \cite{FK14}, starting from the symbol of linearisation 
of equation  (\ref{pdBKP}). Here we do it directly, using the map (\ref{map}).
\subsection*{dBKP equation as a reduction of the MS system}
It is possible to get a condition (\ref{red}) starting from the reduction of the MS hierarchy,
which is characterised by the existence of wave function of adjoint linear operators
of the hierarchy with special analytic properties (with respect to the spectral variable).
The technique to construct this type of reductions was developed in \cite{LVB10}.
Here we will do an elementary derivation on the level of Lax operator for the MS system.

First we introduce formally adjoint linear operators, 
defined by the rule $(u\p)^*=-\p u$ (for all partial derivatives),
\beaa
&&
-X^*=X + \diverg X,
\eeaa
for the Lax operator of the MS system (\ref{MSLax}) we get
\bea
&&
-X_1^*=\partial_y-(\lambda-v_{x})\partial_x + u_{x}\partial_\lambda + v_{xx}= X_1 + v_{xx}
\label{adL}
\eea
We should emphasize that adjoint vector fields in general are not vector fields and contain an extra term
without derivative, equal to divergence of vector field, for zero divergence vector fields are
(anti) self-adjoint. However, the commutation of adjoint vector fields gives the same
compatibility conditions.

Let us suggest that adjoint Lax operator (\ref{adL}) possesses a
wave function of the form 
\bea
\wt \psi= (\lambda- g)^\alpha,\quad X_1^*\wt \psi=0,
\label{red11}
\eea
where $g$ is a function of times. 
This condition  is compatible with the dynamics of the hierarchy and defines a reduction,
see \cite{LVB10}.
The form of this wave function 
can be found considering the map 
from the dispersionless BKP equation to the MS system, we will skip the details.
For the logarithm of the wave function we have an equation
\beaa
X_1 \ln \wt \psi + v_{xx}=
(\partial_y-(\lambda-v_{x})\partial_x + u_{x}\partial_\lambda)\ln \wt \psi + v_{xx}=0,
\eeaa
and, substituting $\wt \psi= (\lambda- g)^\alpha$, we get
\bea
&&
u_x=g_y + g_xv_x -g g_x,
\nn \\
&&
v_{x}=-\alpha g,
\label{red0}
\eea
implying a condition
\bea
u=-\alpha^{-1} v_{y}- \frac{1}{2}(\alpha^{-2}+\alpha^{-1})v_x^2.
\label{red1}
\eea
For $\alpha=-\frac{1}{2}$ this condition coincides with condition (\ref{red})
and MS system (\ref{MSeq}) reduces to potential dispersionless BKP equation (\ref{pdBKP})
($f=v$).  
For $\alpha=0$ relations (\ref{red0}) imply that $v=0$, and we obtain the dKP equation
for the function $u$, $u_x=g_y  - g g_x$.
For general $\alpha$ the MS system reduces to the equation
\bea
v_{xt} = v_{yy}-
\alpha^{-1}(v_{y}+\frac{1}{2}(\alpha^{-1}+1)v_x^2)v_{xx}+v_xv_{xy}-v_{xx}v_y.
\label{alphaeq}
\eea
An interesting special case  corresponds to $\alpha=-1$, then condition 
(\ref{red1})  takes the form $u=v_{y}$ and the MS system reduces to the equation
\bea
v_{xt} = v_{yy}+v_x v_{xy}.
\label{alphaeq1}
\eea
The Einstein-Weyl structure for equation (\ref{alphaeq})
is obtained by the substitution of expression for $u$ (\ref{red1}) 
to the MS system Einstein-Weyl structure (\ref{metricMS}),
\begin{gather*}
\begin{split}
g &= -(dy + v_x dt )^2 +4(dx - 
(\alpha^{-1}+1)( v_{y}+\frac{1}{2\alpha}v_x^2)dt )
dt,
\\
\omega &= v_{xx} dy
+(2(1+2\alpha^{-1})v_{xy} 
+(1+4(\alpha^{-2}+\alpha^{-1}))v_xv_{xx})dt,
\end{split}
\end{gather*}
for equation (\ref{alphaeq1}) it reduces to
\begin{gather*}
\begin{split}
g &= -(dy + v_x dt )^2 +4(dx + dt ) dt,
\\
\omega &= v_{xx} dy+(- 2v_{xy} +v_xv_{xx})dt,
\end{split}
\end{gather*}

It is natural to suggest that, similar to dBKP case, equation
(\ref{alphaeq}) for arbitrary $\alpha$ could be obtained from some Hamiltonian
Lax pair. And it is indeed so!
Let us consider 
dBKP type Lax pair (\ref{HJ0}) with the Hamiltonians
\bea
&&
H_1=\frac{1}{1+\beta}p^{\beta+1} + f_x p,
\nn\\
&&
H_2=\frac{1}{1+2\beta}p^{2\beta+1}
+ f_x p^{\beta+1} + vp,
\label{H1}
\eea
$\beta=2$ corresponds to the dBKP equation case.
In terms of Hamiltonian vector fields we have
\bea
&&
V_1=\p_y - (p^\beta+f_x) \p_x + f_{xx}p\p_p,
\nn\\
&&
V_2=\p_t - (p^{2\beta} + (\beta+1) f_{x}p^\beta + v) \p_x,
+( f_{xx}p^\beta + v_x)p\p_p.
\label{H2}
\eea
Similar to dBKP equation Lax pair, we will transform this Lax pair
to obtain the Lax pair of the MS type.
The first step is to perform a transformation $\mu=p^{\beta}$,
\beaa
&&
V_1=\p_y - (\mu+f_x) \p_x +\beta f_{xx}\mu\p_\mu,
\\
&&
V_2=\p_t - (\mu^{2} +(\beta+1) f_{x}\mu + v) \p_x,
+\beta( f_{xx}\mu + v_x)\mu\p_\mu.
\eeaa
The second step is a transformation $\lambda=\mu + \beta f_x$,
\beaa
&&
V_1=\p_y - (\lambda+(1-\beta)f_{x}) \p_x 
+(\beta f_{xy} - \beta f_x  f_{xx})\p_\mu.
\eeaa
Comparing  with the MS Lax operator (\ref{MSLax}),
we get 
\beaa
&&
u_x=\beta f_{xy} - \beta f_x  f_{xx},
\nn\\
&&
v=(\beta-1)f.
\eeaa
After the identification $g=\beta f_x$, $\beta^{-1}=\alpha +1$,  this transformation
coincides with expressions (\ref{red0}). Thus, equation (\ref{alphaeq})  can be obtained from the 
Hamiltonian Lax pair (\ref{H1}), (\ref{H2}), substituting  $\beta^{-1}=\alpha +1$, $v=\alpha f$.
\subsection*{BMS system}
The symmetry of vector fields $V(-p)=V(p)$  (\ref{vfdBKP}) characterising
dispersionless BKP hierarchy in the framework of dKP hierarchy can be extended to the
Manakov-Santini hierarchy. The Lax pair for the first equation of the hierarchy
(the BMS system) reads
\bea
&&
V_1=\p_y - (p^2 - v_x) \p_x +  u_x p\p_p,
\nn\\
&&
V_2=\p_t - (p^4 + (2u - v_x)p^2 + w_1) \p_x
+ ( u_x p^2 + w_2)p\p_p,
\label{LaxBMS}
\eea
Compatibility conditions imply that $w_2=u_y + (u^2)_x$, and the BMS system can be written in the form
\bea
&&
v_{xt}=v_{xx} w_1 - v {w_1}_x - {w_1}_y,
\nn\\
&&
u_{xt}= u_{yy} + (u^2)_{xy} + u_{xx} w_1
+v u_{xy} + v (u^2)_{xx},
\label{BMS}
\\
&&
{w_1}_x=-2u v_{xx} - v_{xy}.
\nn
\eea
For $u=0$, corresponding to linearly degenerate case, when vector fields
in the Lax pair
do not include the derivative over spectral parameter,
BMS system (\ref{BMS}) reduces to the equation
\beaa
v_{xt}=v_x v_{xy}-v_{xx}{v_y}+v_{yy},
\eeaa
coinciding with the linearly degenerate reduction of MS system
(\ref{MSeq}). It is not unexpected, because in linearly degenerate case we have
a freedom of arbitrary change of the spectral variable
(independent of times).

Hamiltonian reduction corresponds to $u=-v_x$, and system 
 (\ref{BMS}) reduces to potential dBKP equation (\ref{pdBKP}) for the function $f=-v$.
 
 The condition $v=0$ is evidently also a reduction of system  (\ref{BMS}),
 it leads to the equation
 \beaa
 u_{xt}= u_{yy} + (u^2)_{xy} ,
 \eeaa
which coincides with equation (\ref{alphaeq1}) after the identification $v_x=2 u$.

It is rather suprising that, following the steps of transformation of the
dBKP Lax pair to the MS type Lax pair described above,
we are able to define a map (Miura transformation) from solutions
of the BMS system to solutions of the MS system,
thus defining the Einstein-Weyl structure corresponding to the
BMS system.

Performing a transformation $\lambda=p^2+2u$
in the Lax pair (\ref{LaxBMS}), we obtain
a Lax pair of MS type with a Lax operator
\beaa
&&
V_1=\p_y - (\lambda -2u -v_x) \p_x   
+2(u_y +v_x u_x)\p_\lambda.
\eeaa
Comparing this Lax operator to the MS Lax operator (\ref{MSLax}),
we obtain a transformation of solutions of the BMS system
to the solutions of MS system,
\beaa
&&
2u +v_x\rightarrow v_x,
\\
&&
2(u_y +v_x u_x)\rightarrow u_x.
\eeaa
Substituting this transformation to Einstein-Weyl structure
(\ref{metricMS}), we obtain the Einstein-Weyl structure
corresponding to  BMS system (\ref{BMS}),
\begin{gather*}
\begin{split}
g &= -(dy + (2u +v_x )dt )^2 +4(dx + (2\p_x^{-1}(v_xu_x)-v_y ) dt ) dt,
\\
\omega &= (2u_x+v_{xx}) dy+(-8v_x u_x -4 u_y + 2v_{xy} +
\frac{1}{2}((2u +v_x)^2)_x)dt,
\end{split}
\end{gather*}

\subsection*{Hydrodynamic type reductions of the MS system}
Let us consider a multicomponent generalisation of reduction (\ref{red11})
$\wt \psi=\prod (\lambda - g^i)^{\alpha_i}$ 
and complement it with a standard waterbag ansatz \cite{BK04} for 
the wave functions of MS linear operators (\ref{MSLax}) 
$\psi=\lambda+ \sum\gamma_j \ln (\lambda - f^j)$. 
Some special examples of related reductions were considered 
in \cite{Chan10}.
These reductions 
lead to (1+1)-dimensional  systems of hydrodynamic type
defining the dynamics with respect to $y$,
\beaa
&&
g^i_y - (g^i-v_x ) g^i_x - u_x=0,\\
&&
f^j_y - (f^j- v_x) f^j_x - u_x=0,\\
&&
v_x=-\sum\alpha_{i}g^i,\quad
u=-\sum\gamma_{j}f^j
\eeaa
and with respect to $t$,
\beaa
&&
g^i_t -  ((g^i)^2 -v_x g^i + u - v_y) g^i_x -(g^i u_x + u_y)=0,\\
&&
f^j_t - ((g^i)^2 -v_x g^i + u - v_y) g^i_x -(g^i u_x + u_y)=0,\\
&&
\p_x v_y=-\p_y \sum\alpha_{i}g^i =-\sum\alpha_{i}((g^i-v_x ) g^i_x + u_x),\\
&&
u_y=-\sum\gamma_{j} ( (f^j- v_x) f^j_x + u_x).
\eeaa
These (1+1)-dimensional systems are compatible and their common solution represents a 
solution of the MS system.
\section*{Acknowledgements}
This research was performed in the framework of State Assignment of Ministry of Science and Higher Education of the Russian Federation, topic 0029-2021-0004 (Quantum field
theory).


\begin{thebibliography}{99}
\bibitem{Takasaki} K. Takasaki.
Quasi-classical limit of BKP hierarchy and $W$-infinity symmetries.
\textit{Lett. Math. Phys} \textbf{28}, 177--185 (1993)

\bibitem{BK2005}L.V. Bogdanov and B.G. Konopelchenko, On dispersionless BKP hierarchy and its reductions, J. Nonlinear Math. Phys., 12, Suppl.1, 64-73 (2005)

\bibitem{MS06}
S. V. Manakov and P. M. Santini,
The Cauchy problem on the plane for the dispersionless
Kadomtsev-Petviashvili equation,
\textit{JETP Lett.} \textbf{83} (2006) 462--466.

\bibitem{MS07}
S. V. Manakov and P. M. Santini,
A hierarchy of integrable PDEs in 2+1 dimensions associated with
2-dimensional vector fields,
\textit{Theor. Math. Phys.} \textbf{152} (2007) 1004--1011.

\bibitem{DFK14}M. Dunajski, E.V. Ferapontov and 
B. Kruglikov, On the Einstein-Weyl 
and conformal self-duality equations,
Journal of Mathematical Physics 56(8), 083501 (2015).

\bibitem{PT93} H. Pedersen and K.P. Tod, 
Three Dimensional Einstein-Weyl Geometry,
Advances in Mathematics,
97 (1), 74--109 (1993)

\bibitem{FK14}
E.V. Ferapontov and B.S. Kruglikov,
Dispersionless integrable systems in 3D and Einstein-Weyl geometry,
J. Differential Geom.
Volume 97, Number 2 (2014), 215-254.

\bibitem{Hitchin} N.J. Hitchin,  
{Complex manifolds and Einstein's equations}, 
Twistor geometry and nonlinear systems (Primorsko, 1980), 
73--99, 
Lecture Notes in Math. {\bf 970}, 
Springer, Berlin-New York (1982).

\bibitem{LVB10}L.V. Bogdanov, On a class of reductions of the Manakov-Santini hierarchy connected with the interpolating system, J. Phys. A: Math. Theor. 43, 115206 (2010)

\bibitem{BK04}L.V. Bogdanov and B.G. Konopelchenko, Symmetry constraints for dispersionless integrable equations and systems of hydrodynamic type, Phys. Lett. A 330(6), 448-459 (2004)

\bibitem{Chan10}Jen-Hsu Chang and Yu-Tung Chen, Hodograph solutions for the Manakov–Santini equation, Journal of Mathematical Physics 51, 042701 (2010)


\end{thebibliography}
\end{document}